\documentclass[aip,rsi,reprint]{revtex4-1}
\usepackage{amssymb}
\usepackage{amsmath}
\usepackage{textcomp}
\usepackage{graphicx} 
\usepackage{epstopdf}
\usepackage{dcolumn}
\usepackage{bm}
\usepackage{soul}

\def\um{\upmu\mbox{m}}

\def\Wcm2{\mbox{W cm}^{-2}}
\def\Wcmum2{\mbox{Wcm}^{-2}\upmu\mbox{m}^{2}}

\def\cm3{\mbox{cm}^{-3}}

\usepackage{hyperref}
\hypersetup{colorlinks=true,linkcolor=red,citecolor=red,urlcolor=red}
\usepackage{upgreek}

\begin{document}
\title{Characterisation of deuterium spectra from laser driven multi-species sources by employing differentially filtered image plate detectors in Thomson spectrometers}

\author{A.~Alejo}
\affiliation{Centre for Plasma Physics, School of Mathematics and Physics, Queen's University Belfast, BT7 1NN, UK}

\author{S.~Kar}\email{s.kar@qub.ac.uk}
\affiliation{Centre for Plasma Physics, School of Mathematics and Physics, Queen's University Belfast, BT7 1NN, UK}

\author{H.~Ahmed}
\affiliation{Centre for Plasma Physics, School of Mathematics and Physics, Queen's University Belfast, BT7 1NN, UK}

\author{A.G.~Krygier}
\affiliation{Department of Physics, The Ohio State University, Columbus, Ohio 43210, USA}

\author{D.~Doria}
\affiliation{Centre for Plasma Physics, School of Mathematics and Physics, Queen's University Belfast, BT7 1NN, UK}
\author{R.~Clarke}

\affiliation{Central Laser Facility, Rutherford Appleton Laboratory, Didcot, Oxfordshire, OX11 0QX, UK}
\author{J.~Fernandez}

\affiliation{Instituto de Fusi\'on Nuclear, Universidad Polit\'ecnica de Madrid, 28006 Madrid, Spain}
\affiliation{Central Laser Facility, Rutherford Appleton Laboratory, Didcot, Oxfordshire, OX11 0QX, UK}

\author{R.R.~Freeman}
\affiliation{Department of Physics, The Ohio State University, Columbus, Ohio 43210, USA}

\author{J.~Fuchs}
\affiliation{LULI, Ecole Polytechnique, CNRS, Route de Saclay,
91128 Palaiseau Cedex,France}

\author{A.~Green}
\affiliation{Centre for Plasma Physics, School of Mathematics and Physics, Queen's University Belfast, BT7 1NN, UK}

\author{J.S.~Green}
\affiliation{Central Laser Facility, Rutherford Appleton Laboratory, Didcot, Oxfordshire, OX11 0QX, UK}

\author{D.~Jung}
\affiliation{Centre for Plasma Physics, School of Mathematics and Physics, Queen's University Belfast, BT7 1NN, UK}

\author{A. Kleinschmidt}
\affiliation{Institut f\"ur Kernphysik, Technische Universit\"at Darmstadt, Schlo{\ss}gartenstrasse 9, D-64289 Darmstadt, Germany}

\author{C.L.S. Lewis}
\affiliation{Centre for Plasma Physics, School of Mathematics and Physics, Queen's University Belfast, BT7 1NN, UK}

\author{J.T.~Morrison}
\affiliation{Propulsion Systems Directorate, Air Force Research Lab, Wright Patterson Air Force Base, Ohio 45433, USA}

\author{Z.~Najmudin}
\affiliation{Blackett Laboratory, Department of Physics, Imperial College, London SW7 2AZ, UK}

\author{H.~Nakamura}
\affiliation{Blackett Laboratory, Department of Physics, Imperial College, London SW7 2AZ, UK}

\author{G.~Nersisyan}
\affiliation{Centre for Plasma Physics, School of Mathematics and Physics, Queen's University Belfast, BT7 1NN, UK}

\author{P.~Norreys}
\affiliation{Central Laser Facility, Rutherford Appleton Laboratory, Didcot, Oxfordshire, OX11 0QX, UK}
\affiliation{Department of Physics, University of Oxford, Oxford, OX1 3PU, UK}

\author{M.~Notley}
\affiliation{Central Laser Facility, Rutherford Appleton Laboratory, Didcot, Oxfordshire, OX11 0QX, UK}

\author{M.~Oliver}
\affiliation{Department of Physics, University of Oxford, Oxford, OX1 3PU, UK}

\author{M.~Roth}
\affiliation{Institut f\"ur Kernphysik, Technische Universit\"at Darmstadt, Schlo{\ss}gartenstrasse 9, D-64289 Darmstadt, Germany}

\author{J.A.~Ruiz}
\affiliation{Instituto de Fusi\'on Nuclear, Universidad Polit\'ecnica de Madrid, 28006 Madrid, Spain}

\author{L.~Vassura}
\affiliation{LULI, Ecole Polytechnique, CNRS, Route de Saclay,
91128 Palaiseau Cedex,France}

\author{M.~Zepf}
\affiliation{Centre for Plasma Physics, School of Mathematics and Physics, Queen's University Belfast, BT7 1NN, UK}
\affiliation{Helmholtz Institut Jena, D-07743 Jena, Germany}

\author{M.~Borghesi}
\affiliation{Centre for Plasma Physics, School of Mathematics and Physics, Queen's University Belfast, BT7 1NN, UK}
\affiliation{Institute of Physics of the ASCR, ELI-Beamlines project, Na Slovance 2, 18221 Prague, Czech Republic}

\date{\today}

\begin{abstract}
A novel method for characterising the full spectrum of deuteron ions emitted by laser driven multi-species ion sources is discussed. The procedure is based on using differential filtering over the detector of a Thompson parabola ion spectrometer, which enables discrimination of deuterium ions from heavier ion species with the same charge-to-mass ratio (such as C$^{6+}$, O$^{8+}$, etc.). Commonly used Fuji Image plates were used as detectors in the spectrometer, whose absolute response to deuterium ions over a wide range of energies was calibrated by using slotted CR-39 nuclear track detectors. A typical deuterium ion spectrum diagnosed in a recent experimental campaign is presented, which was produced from a thin deuterated plastic foil target irradiated by a high power laser.
\end{abstract}

\pacs {} %

\maketitle

\section{Introduction}

Techniques for accelerating ions employing high power lasers are currently attracting considerable interest, thanks to the rapid progress in laser technologies over the past couple of decades. With currently accessible intensities (in excess of $10^{20}~\Wcm2$) of short pulse lasers, ions can be accelerated to high energies (up to several tens of MeV/nucleon) with promising beam parameters for several potential applications in science, industry and healthcare~\cite{ion_review-1,ion_review-2,ion_review-3}. A number of acceleration mechanisms, such as Target Normal Sheath Acceleration (TNSA)~\cite{TNSA_2,TNSA_3}, Radiation Pressure Acceleration (RPA)~\cite{RPA_2,RPA_3,RPA_4,RPA_7}, Break-Out Afterburner (BOA)~\cite{BOA_2,BOA_5}, are objects of intense investigation both experimentally and theoretically in order to improve the beam parameters. 

Laser accelerated ion beams from solid targets are typically multi-species, either due to the chemical composition of the target material or due to the layer of contamination over the target (typically composed of water vapour and hydrocarbons)\cite{gitomer1986fast,wilks2001energetic}. One of the diagnostics commonly used to characterize these beams is the Thomson Parabola Spectrometer (TPS)\cite{Thomson1913,Gwynne2014}, which has the unique ability to energy resolve the ion spectra while discriminating ions with different charge-to-mass ratio.
\begin{figure}
\begin{center}
\includegraphics[width=0.48\textwidth]{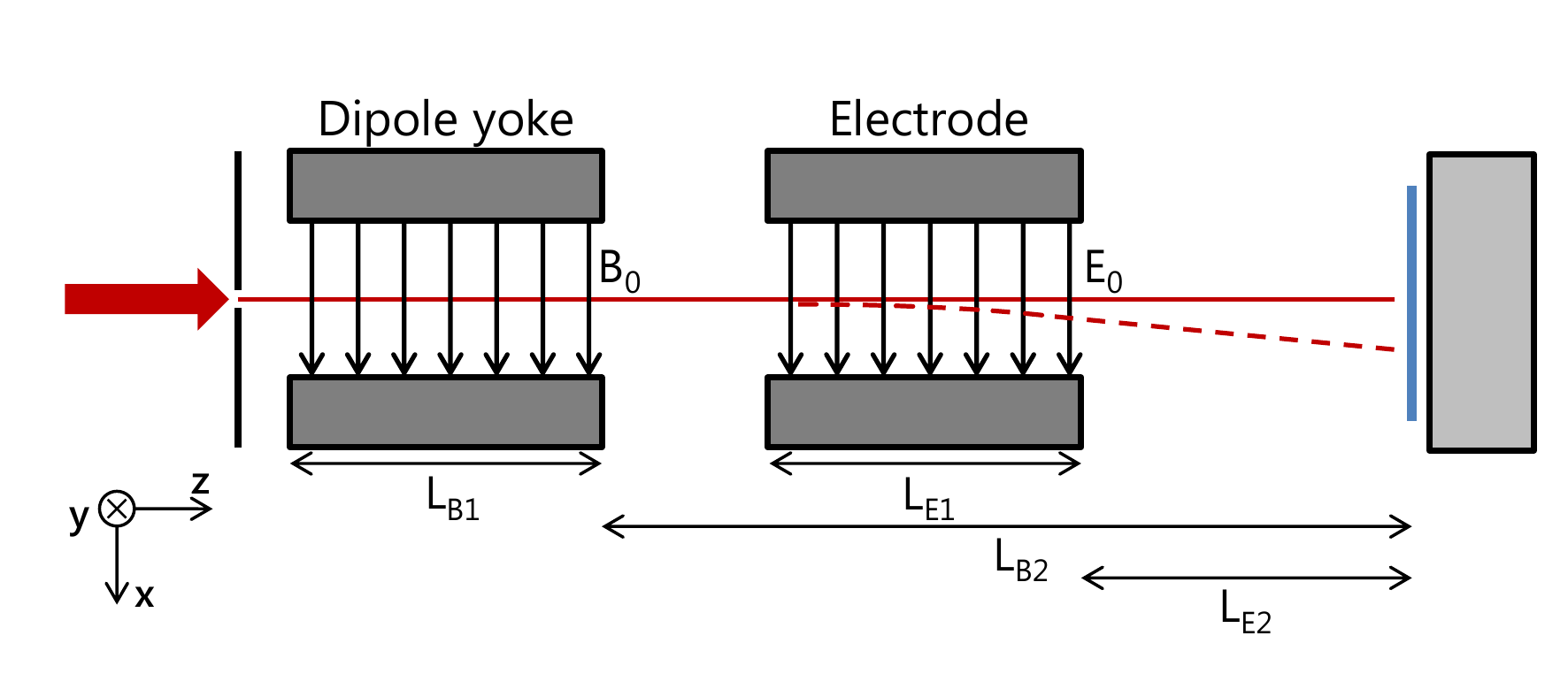}
\end{center}
\caption{\textbf{(a)} Schematic of a typical TPS with regions of static magnetic and electric fields for energy dispersion and separation of ion tracks with different $q/m$ ratios, respectively, as shown in \textbf{(b)}. A set of filters were used over the detector in order to discriminate lighter ions from the overlapping heavier species with the same $q/m$ ratio.}
\label{fig:TPIS_Setup}
\end{figure}

In a typical TPS (Fig.~\ref{fig:TPIS_Setup}), a pencil beam of ions, selected by the pinhole located at its entrance, travels through regions of parallel magnetic and electric fields, applied transversely to the beam axis, before reaching the detector plane. With reference to fig.~\ref{fig:TPIS_Setup}, the magnetic field determines the position of the ions along the $y$-axis, depending on their energy per nucleon ($W$)~(or longitudinal velocity ($v_{z}$)), while the electric field deflects the ions along the the $x$ axis according to their charge($q$) to mass($m$) ratio. Assuming that the inhomogeneities of the fields are negligible and that the ion energies are not relativistic, particles with given $q/m$ will lay on a parabolic trace on the detector plane described by the expression
\begin{equation}\label{Eq:TPISParabola}
y^2=\frac{q}{m}\frac{B_0^2}{E_0}
\frac{{L_{B1}}^2\left(L_{B1}/2+L_{B2}\right)^2}
{L_{E1}\left(L_{E1}/2+L_{E2}\right)}\cdot x
\end{equation}

where $E_0$ and $B_0$ are the electric and magnetic fields, $L_{B1}$, $L_{B2}$, $L_{E1}$ and $L_{E2}$ are the dimensions of relevant sections of the TPS as labelled in Fig.~\ref{fig:TPIS_Setup}. Since the locus of the parabolic ion traces produced by the TPS is a function of the charge-to-mass ratio ($q/m$) of the ion species, the traces of the species with the same $q/m$ will overlap at the detector plane, preventing their spectra to be characterised. This is a fundamental limitation of a TPS when employed in high intensity laser plasma experiments, where several ion species around the interaction region are simultaneously accelerated to high energies~\cite{heavy_ion_McKenna_PPCF_2007,RPA_7,BOA_5}. 
Examples of fully ionized ions with the same $q/m$ ratios which are typically observed in laser-plasma experiments are  D$^+$, C$^{6+}$ and O$^{8+}$ ($q/m$=1/2) and C$^{3+}$ , O$^{4+}$ ($q/m$=1/4). 
Characterization of the overlapping ion species individually is in many cases important not only in order to understand the underlying acceleration mechanism, but also to facilitate the analysis of any data obtained from secondary processes involving these ion species. This is the case for laser driven deuterium ions (D$^+$)\cite{morrison2012selective,maksimchuk2013dominant}, which are being used to explore the potential of compact, laser-based sources of neutrons\cite{NeutronRoth2013,willingale2011comparison,zulick2013energetic} with possible applications in science~\cite{Higginson_NRS}, industry~\cite{perkins2000investigation, Nakai_industry}, security~\cite{Loveman1995765} or healthcare~\cite{Wittig_BNCT}. The most commonly used targets to produce energetic beams of D$^+$ ions are deuterated plastic foils, and a direct measurement of the D$^+$ ion spectra with a TPS is non-trivial due to the overlapping C$^{6+}$ and O$^{8+}$ ions from the target bulk and contaminant layers.  

Discrimination of D$^+$ ions from the overlapping heavier ions can be achieved by filtering out the heavier ions before reaching the detector, for instance, by using a metal foil of appropriate thickness which can preferentially stop heavier ions. This is possible as, along a ion trace corresponding to a given value of $q/m$, all the ions will have the same energy per nucleon at any given point of the trace. In principle, heavier species can therefore be prevented from reaching the detector by exploiting differences in stopping power of the different ions, if a suitable filter can be selected. However, using a metal foil with a single, fixed thickness in front of the detector may not be sufficient to recover the full D$^+$ spectrum, depending on the energy range covered by the TPS. Therefore, a possible solution is to filter differentially different sections of the ion trace on the detector. 

In this paper, we report on the characterisation of the complete spectrum of D$^+$ ions generated from a laser solid interaction by employing a differential filter (DF) over FUJI image plate detectors~\cite{FujifilmWeb}, which are commonly used in such experiments. The paper is organised as follows. After a brief description of the experimental setup in section \ref{sec:ExperimentalSetup}, section \ref{sec:Filtering} describes the concept of differential filtering and the design used in the experiment. Section \ref{sec:Calibration} reports the absolute calibration of the image plate detector for D$^+$ ions over a wide range of energies, which was obtained in a single shot by using slotted CR-39 nuclear track detectors placed over the image plates. Finally, a typical D$^+$ spectrum is shown in section~\ref{sec:Application}, which was obtained by the analysis of a typical TPS data from the experiment.

\section{Experimental Setup}\label{sec:ExperimentalSetup}

The experiment was carried out at the Rutherford Appleton Laboratory (RAL), STFC, UK by employing the petawatt arm of the VULCAN laser system. The data presented in this paper were taken by irradiating the laser onto $10~\um$ thick deuterated plastic (CD) foil targets. Using a $f/3$ off-axis parabolic mirror, the laser was focussed down to $\sim 6~\um$ full width at half maximum spots on the target, delivering peak intensity in excess of $10^{20}~\Wcm2$. The ions accelerated by the interaction of the intense pulse with the CD target were diagnosed by employing several TPSs, at different angles with respect to the laser axis.  

As discussed before, such interaction produces a multi-species ion source of broad energy spectrum. Therefore, the accelerated deuterium ions were discriminated from the heavier ones with the same $q/m$ value (such as C$^{6+}$, O$^{8+}$, etc. present in both the target bulk and contaminant layers) by using differential filtering in front of the detector of each TPS. The DF (as discussed in section~\ref{sec:Filtering}) was attached to a frame which was directly screwed onto the detector drum of the TPS, and served as a clamp to hold the detector plate (see Fig. \ref{fig:TPIS_Setup}). In this configuration the position of the DF always remains fixed with respect to the TPS, which is crucial for filtering ions in specific energy bands. BAS-TR imaging plates (IP)~\cite{FujifilmWeb} were used as detectors. The IPs were wrapped with $6~\um$ thick Al foils to avoid their exposure to ambient light after irradiation.

In the shots taken for image plate calibration, slotted CR-39 track detectors~\cite{fleischer1965CR39} were placed between the DF and the IP. The CR-39s were slotted uniformly along the energy-dispersion axis in order to select small parts of the ion tracks for absolute particle counting. More details about the technique will be discussed in section~\ref{sec:Calibration}. 

\section{Design of a Differential Filter (DF)}\label{sec:Filtering}

As discussed before, tracks of ions with the same $q/m$ value overlap on the detector, where the particles at a given point of the track share the same energy per nucleon. This allows using a metal foil of suitable thickness to stop the heavier ions while allowing the lighter ions to pass through due to the higher stopping power of the heavier ions. However, due to the broad energy spectrum of the ions produced in such experiments, a single filter cannot be used across the full spectrum, as the given filter foil would either be too thin to filter the heavier ions at higher energies or too thick to let lower energy deuteron ions to pass through. The design of a differential filter is based on using several filters of suitable materials and/or thicknesses in different regions of the detector, set along the energy dispersion axis of the detector. The requirement in this case is that for each zone, with boundaries defined by a pair of Y coordinates $(y_1, y_2)$, such a filter should be capable of stopping the heavier ions reaching the detector plane within this region.

\begin{figure}
\includegraphics[width=0.48\textwidth]{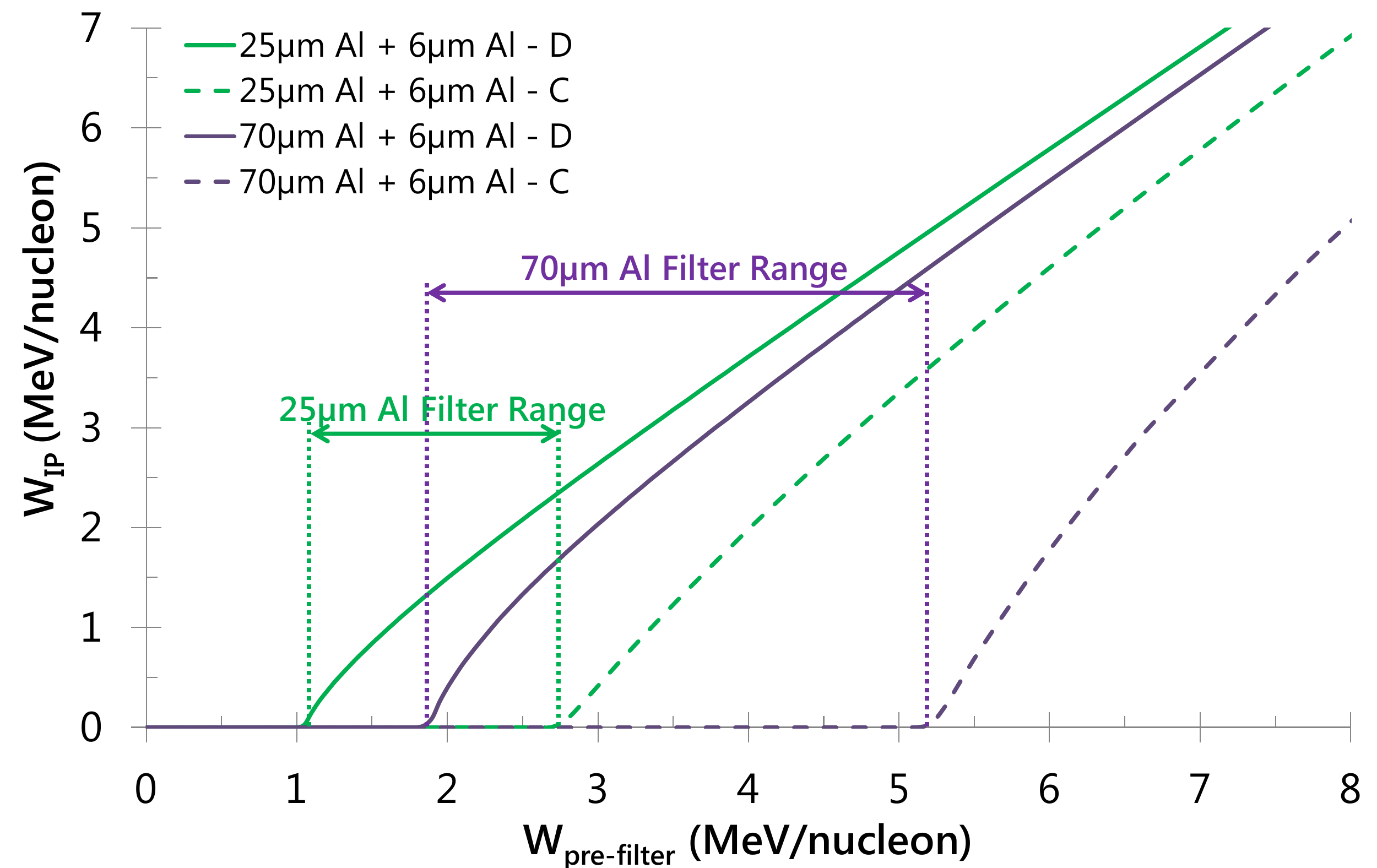}
\caption{Graph showing transmitted vs. incident energies for deuterium (solid) and carbon (dashed) ions for 25~$\um$ (blue) and 70~$\um$ (red) aluminium foils, respectively. The grey blocks represent the regions selected for filtering carbon ions from the low energy part of the deuterium track using the two filters.}
\label{fig:Filtering}
\end{figure}

It is to be noted that a DF has to be designed specifically for a given TPS and for the ion species under consideration, as those are the parameters that defines the energy of the ions to filter at each point of the detector. In this paper we discuss filtering of C$^{6+}$, O$^{8+}$ions in order to obtain a clean track of D$^{+}$ on the detector. In the experiment, the TPS was setup (with $B_0$ = 0.988~T, $E_0$ = 11~kV/cm, $L_{B1}$ = 50~mm, $L_{B2}$ = 407~mm, $L_{E1}$ = 150~mm, $L_{E2}$ = 307~mm) to diagnose D${}^+$ ions from $\sim$1~MeV/nucleon up to several tens of MeV/nucleon, spread over 55~mm along the energy dispersion axis of the detector. Considering the energy dispersion of the D${}^{+}$ and C${}^{6+}$ ions, the 55~mm detection window was divided into 4 regions, where a suitable filter could be found for each region. The filter foils for each region were selected by considering the stopping power of both deuterium and carbon ions, obtained via SRIM simulations~\cite{Ziegler2010}, as shown in Fig.~\ref{fig:Filtering}. 

The set of four filters was selected to cover the entire energy range of D${}^+$ ions detectable by the TPS as listed in Table~\ref{table:Filters}. It is to be noted that the choice of filter materials and thicknesses were based on their availability during the experiment. Other materials of suitable thicknesses can be found for similar performances, however, higher Z and thinner foils would be preferred in order to decrease lateral straggling of ions due to multiple small-angle scattering inside the filters, which may reduce the energy resolution of the diagnostic. The filters can be assigned to energy bins far from their filtering limits, which was the case in our experiment. Overall, slight variations around the thickness and size of the chosen filter is acceptable, making the filter selection and DF deployment easier.

\begin{table}[h!]
\begin{center}
\begin{tabular}{| c | c | c | c |}
\hline 
Material & Thickness & $\mbox{E}_{\mbox{D,min}}$ & $\mbox{E}_{\mbox{D,max}}$\\
 & ($\um$) & (MeV/nucleon) & (MeV/nucleon)\\
\hline
Al & 25 & 1.1 & 2.2\\
Al & 70 & 2.3 & 4.8\\
Fe & 125 & 4.85 & 10.3\\
Cu & 250 & 10.5 & 30.0\\
\hline
\end{tabular}
\end{center}
\caption{The four foils selected for the DF used in the experiment for different energy bins of D${}^+$ ions in order to filter carbon and heavier ions}
\label{table:Filters}
\end{table}

\begin{figure}
\includegraphics[width=0.48\textwidth]{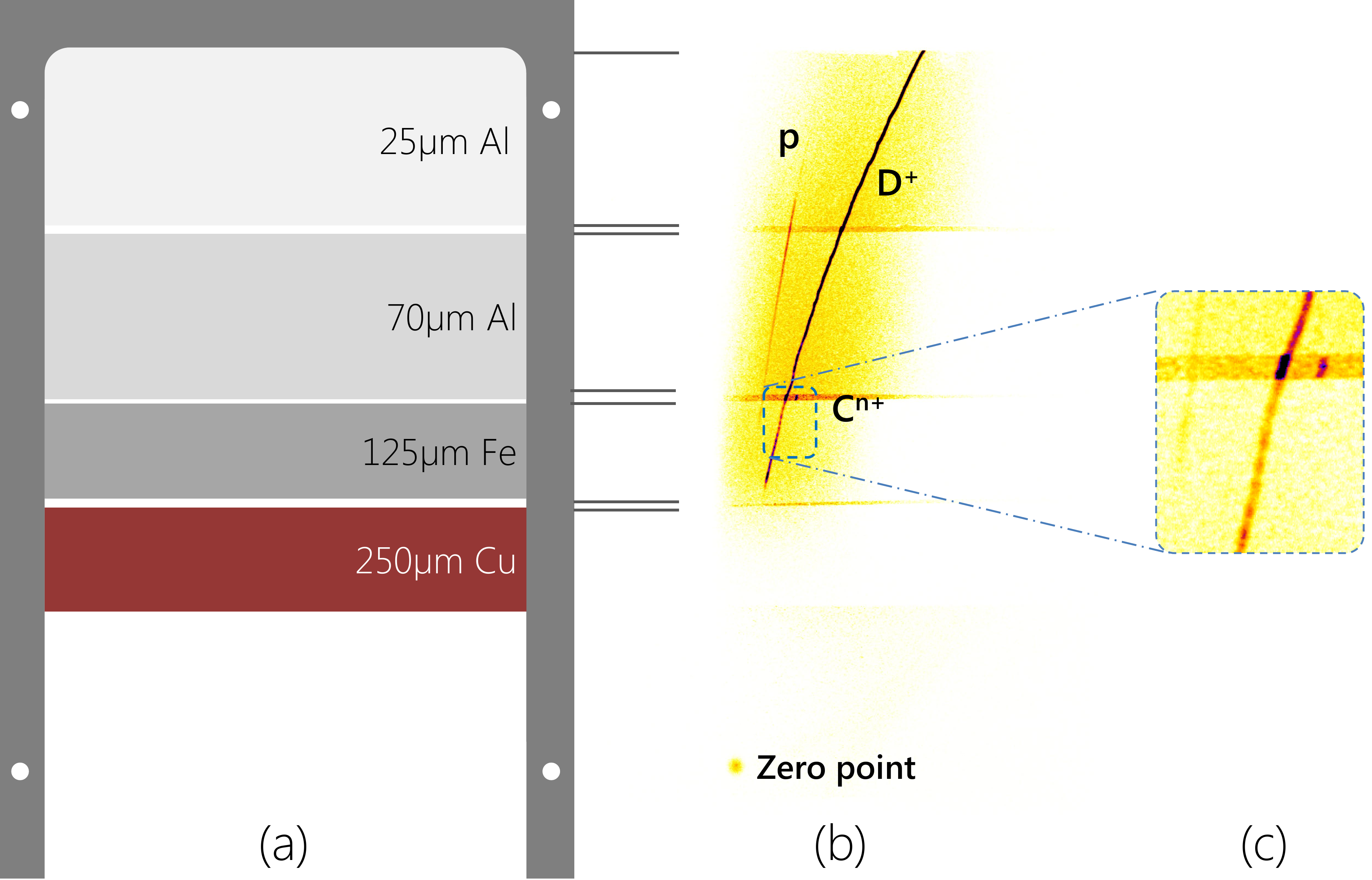}
\caption{\textbf{(a)} Schematic of a DF design used in the experiment. \textbf{(b)} shows typical raw data obtained in the experiment using the DF shown in (a). \textbf{(c)} shows a zoomed-in view of a small region in the data (as marked by dashed square in (b)) highlighting the abrupt increase in signal of "D${}^+$ track" and appearance of other ion species in the gap between two filter foils.}
\label{fig:FilterExample}
\end{figure}

Finally, in order to construct the DF, the energy bounds for each of the energy bins were converted into corresponding y coordinates on the detector by taking into account the energy dispersion of the TPS. A schematic of the DF used in the experiment is shown in the Fig.~\ref{fig:FilterExample}, along with typical data obtained in the campaign.

As can be seen in Table~\ref{table:Filters}, a small gap was kept between the selected energy bins, corresponding to $\sim$1 mm gap between the filters over the IP. The gaps between the filters were made not only to check the filter performance, but also to identify the location of different filters from their shadows in the scanned image of the IP. One can notice the abrupt increase of the signal in the ion trace corresponding to $q/m$=1/2 and the appearance of other ion species tracks (such as C${}^{5+}$, O${}^{7+}$, etc.) in the gap between consecutive filter sheets (Fig.~\ref{fig:FilterExample}c). This is evidence of the effectiveness of the DF design used in the TPS for isolating deuterium ion spectra from the other ion species with the same $q/m$ value.

\section{Calibration of IP response to energetic deuterium ions}\label{sec:Calibration}

Once the carbon and heavier ion species have been filtered from the D${}^+$ track by using the DF, the number of deuterium ions incident on the detector need to be calculated using the detector response (in our case IPs). The IP detectors are based on the principle of photo-stimulated luminescence (PSL)~\cite{FujifilmWeb}. When ionising radiation is incident on the phosphor layer of the IP (usually europium-doped barium fluorohalide phosphor), electrons are excited to higher energy levels in the crystal lattice where they remain in a metastable level until stimulated by a secondary illumination of appropriate wavelength. An accurate theoretical response function of the detector is difficult to obtain due to the complexity of the mechanism of electron excitation and degree of trapping as a function of the type and energy of the incident ionising radiation. For this reason, a direct calibration method was performed to calibrate the IP response for deuterium ions. The calibration curve refers to a relationship between the PSL signal produced in the IP per incident particle versus the energy of the particle at the detector.

\begin{figure}
\includegraphics[width=0.4\textwidth]{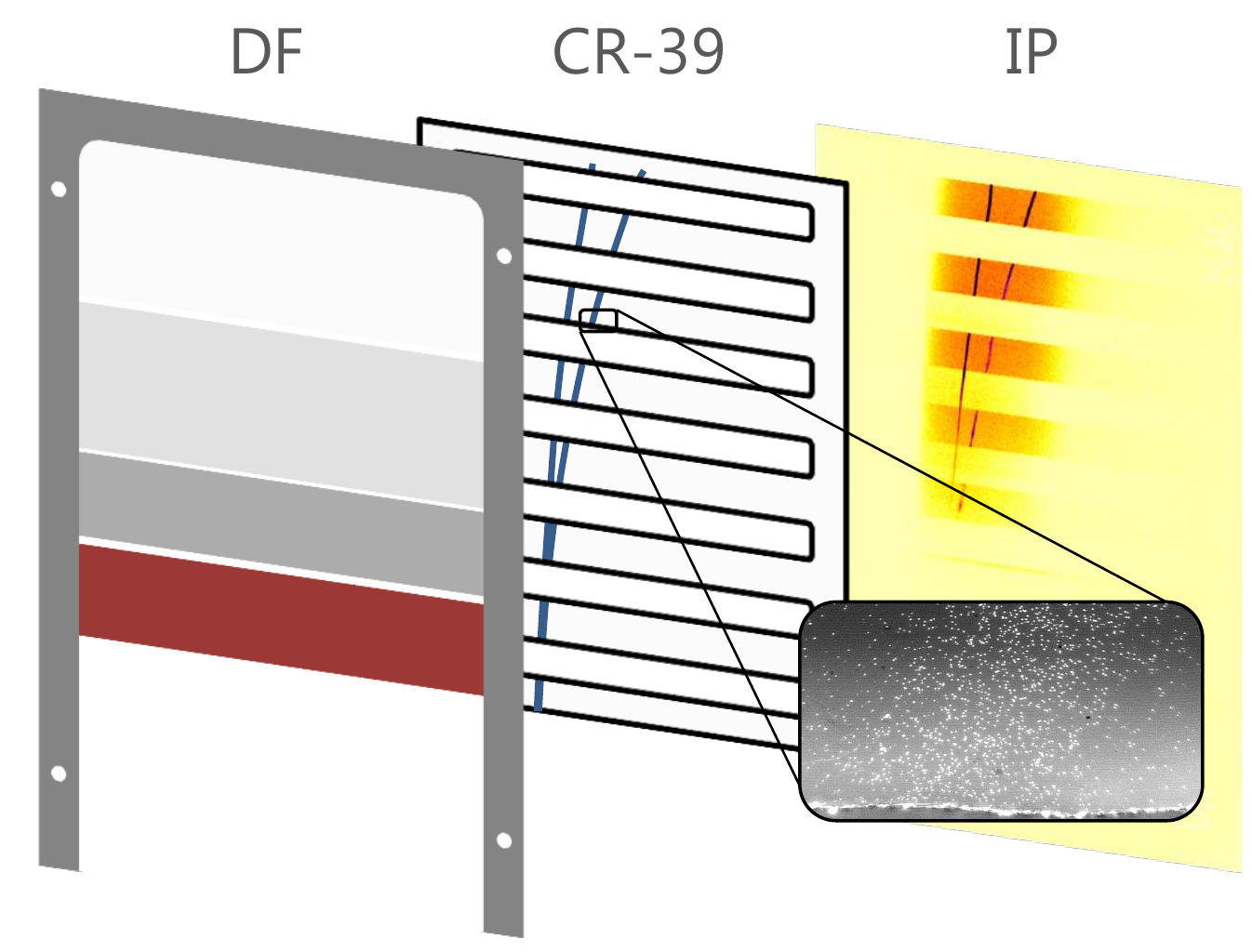}
\caption{Example of the setup used for the BAS-TR IP calibration, showing experimental results from both the CR-39 detector and the image plate.}
\label{fig:Calibration_Setup}
\end{figure}

The calibration method, as used previously by several groups~\cite{mori2006, mancic2008}, involves using CR-39 nuclear track detectors to compare the PSL signal obtained from the IP with the number of tracks produced in the CR-39. This method provides an absolute ion calibration, as each track produced in the CR-39 refers to a single ion. To obtain the IP response for a range of ion energies,  a piece of CR-39 having closely spaced slots carved along the IP energy dispersion axis (as shown in the Fig.~\ref{fig:Calibration_Setup}) was placed in front of the IP~\cite{prasad2010calibration}. The slotted CR-39 was used after the DF, ensuring that only the D$^+$ ions are exposed to the CR-39. Several data sets were collected by employing this setup in different TPS used in the experiment, where each set of CR-39 and IP data provided 4-5 data points towards the calibration curve. The methods employed for obtaining the calibration curve from the experimental data are discussed below.

\subsection{Track counting in CR-39}
In order to develop the ion tracks for counting, the CR-39 pieces were etched in 6M sodium hydroxide (NaOH) solution at a temperature of $85\,^o\mbox{C}$. Etching was done for short periods of time (typically 5-10 mins), depending on the incident flux of particles on the CR-39, to avoid overlapping of ion tracks. In order to facilitate track counting, sets of pictures of the front surface of CR-39 were taken using an optical microscope with a $20\times$ objective. The number of particle tracks at any given location was then counted by using the 'particle analysis' subroutine of ImageJ\cite{ImageJ} software. The accuracy of this method was cross-checked by counting the tracks manually for several cases, which showed differences smaller than $5\%$ in every case. 

\subsection{Estimation of IP signal incorporating PSL fading}

\begin{figure}
\includegraphics[width=0.49\textwidth]{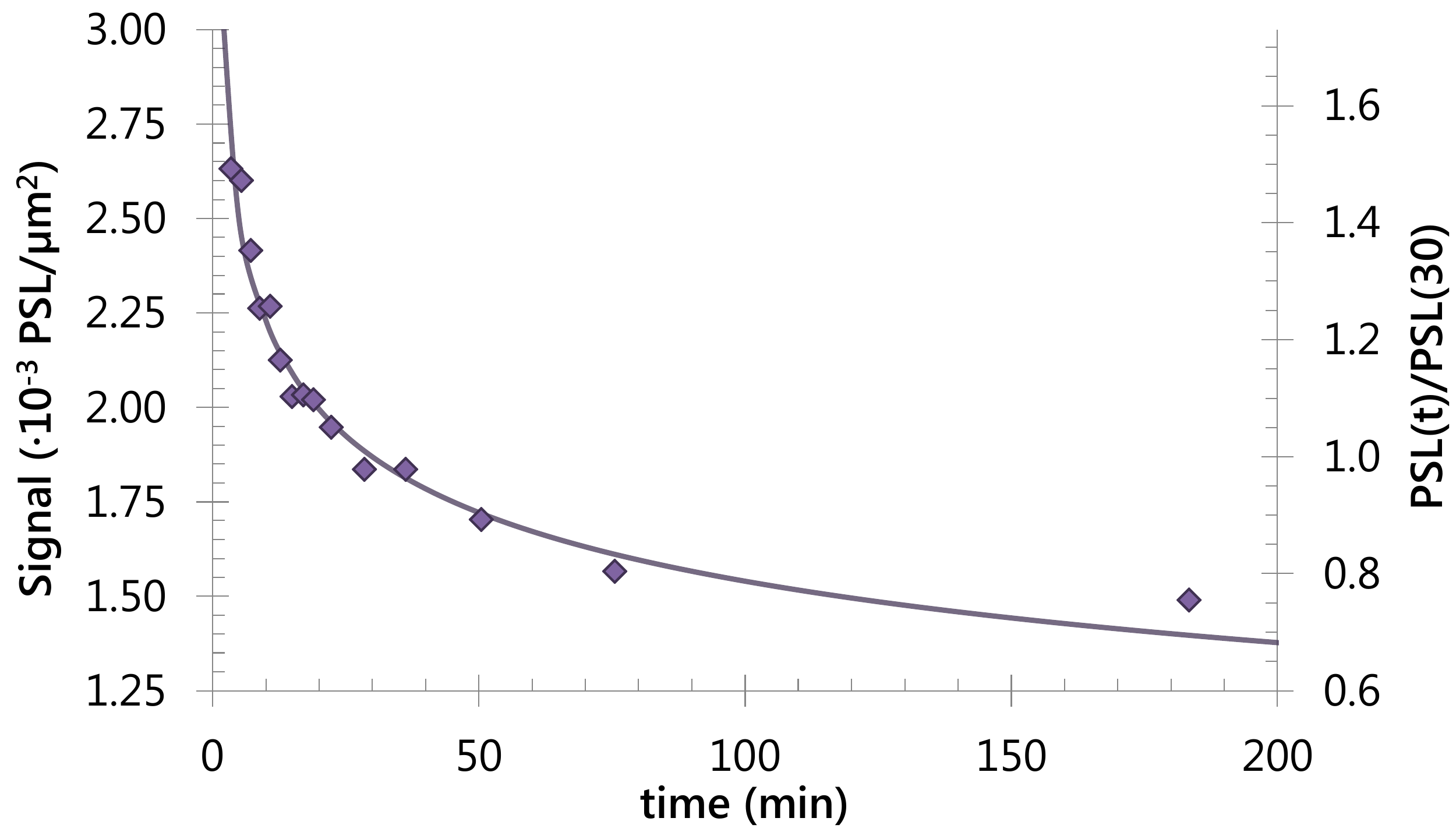}
\caption{Experimentally measured PSL signal per unit area for different delayed scanning times are shown for the BAS-TR IP, while using $25\times 25\um$ scanning resolution. The solid line represents the fading curve, which is the best fit to the data points.}
\label{fig:Fading}
\end{figure}

The IPs were scanned using a commercial IP scanner (Fujifilm FLA-5000~\cite{FujifilmWeb}) with 16-bit dynamic range and  $25~\um~\times~25~\um$ pixel size, which was converted from the scan value (or quantum level (QL)) to PSL using the formula given by the manufacturer~\cite{FujifilmWeb2}
\begin{equation} PSL=\left(\frac{R}{100}\right)^2\times  \frac{4000}{S}\times 10^{L\times \left(\frac{QL}{2^G-1}-\frac{1}{2}\right)}\end{equation}
where $R=25\um$ is the scanning resolution, $S=5000$ is the sensitivity of the scanner, $L=5$ is the latitude or level and $G=16$ is the bit value of the dynamic range of the image. 

Due to the spontaneous decay of excited electrons from their metastable state, energy stored in the IP decays over time after the irradiation. As shown by several authors~\cite{mancic2008,paterson2008,tanaka2005, ohuchi2002} the fading of the IP signal can be describe by different rates of decay, form a very fast decay over several tens of minutes after the irradiation to a significantly slower rate of decay lasting for days. The fading has usually been investigated by exposing the IP to radioactive sources (e.g. alpha, gamma) for a period of time ranging from few minutes to tens of minutes and scanning it after the exposure. Although this method is rigorous in describing the slow rate of decay, it does not allow sufficient accuracy for the measuremnt of the decay rate promptly after the irradiation. Moreover, it is likely that the fading rate close to the time of irradiation can vary significantly depending on several factors such as the PSL signal strength at the time of exposure, duration of ionising radiation source, type of image plate, etc.

In order to overcome these factors, the decay characteristic of the BAS-TR image plate was studied experimentally by exposing the IP to a laser driven pulsed X-ray source of ns burst duration, employing TARANIS laser facility at the Queen's University Belfast, UK. Small pieces (few mm $\times$ few mm) of BAS-TR IP were placed at a given distance from a laser target in order to illuminate the IPs to a spatially homogeneous X-ray beam, while producing equivalent level of PSL as obtained by the ions in our experiment. After the exposure, the image plates were scanned at different times ranging from few minutes to few hours after the exposure. As shown in Fig.~\ref{fig:Fading}, PSL signal suffers a rapid decay in the first half hour after the exposure, followed by a significantly slower decay over several hours. Although the trend of the fading curve broadly matches the measurements taken by other groups (see the references ~\cite{mancic2008,paterson2008,tanaka2005}), the agreement is best for long fading times of the order of hours.

In our case the TPS were placed inside the main interaction chamber, which was kept under vacuum during the interaction. This caused a significant delay in scanning the IPs after the irradiation, which ranged between 30 minutes up to 2 hours for different shots. Therefore, instead of obtaining a calibration curve in terms of PSL per ion at the time of exposure (i.e. t = 0), it was decided to set the reference time at $t = 30$ minutes after irradiation, which is a typical time required for retrieving the IPs from the vacuum interaction chamber for scanning. Based on the data shown in the Fig.~\ref{fig:Fading}, the PSL signal at the said reference point of time, i.e. 30 min. after exposure, was estimated by the empirical formula

\begin{equation}\label{eq:Fading}
\mbox{PSL}_{30}=\left(\frac{30}{t}\right)^{-0.161}\mbox{PSL}(t)
\end{equation}
where PSL(t) represents the measured PSL signal obtained from the IP scanned `$t$' minutes after the irradiation. 

\subsection{Energy transmission}
As mentioned earlier, a calibration curve refers to the signal produced per ion as a function of ion energy per nucleon reaching the IP (W$_{\mbox{IP}}$). Due to the use of the DF before the detectors in our case, the energy of deuterium ions incident on the IP/CR-39 at any given point of the ion track was different from that defined by the TPS energy dispersion. Therefore, for each data point obtained for the calibration curve, the energy of the D$^+$ ion on the image plate, after passing through the DF, was calculated by using SRIM simulation~\cite{Ziegler2010}, as shown in Fig.~\ref{fig:Filtering} for the first two filter layers. 

\begin{figure}[h]
\includegraphics[width=0.49\textwidth]{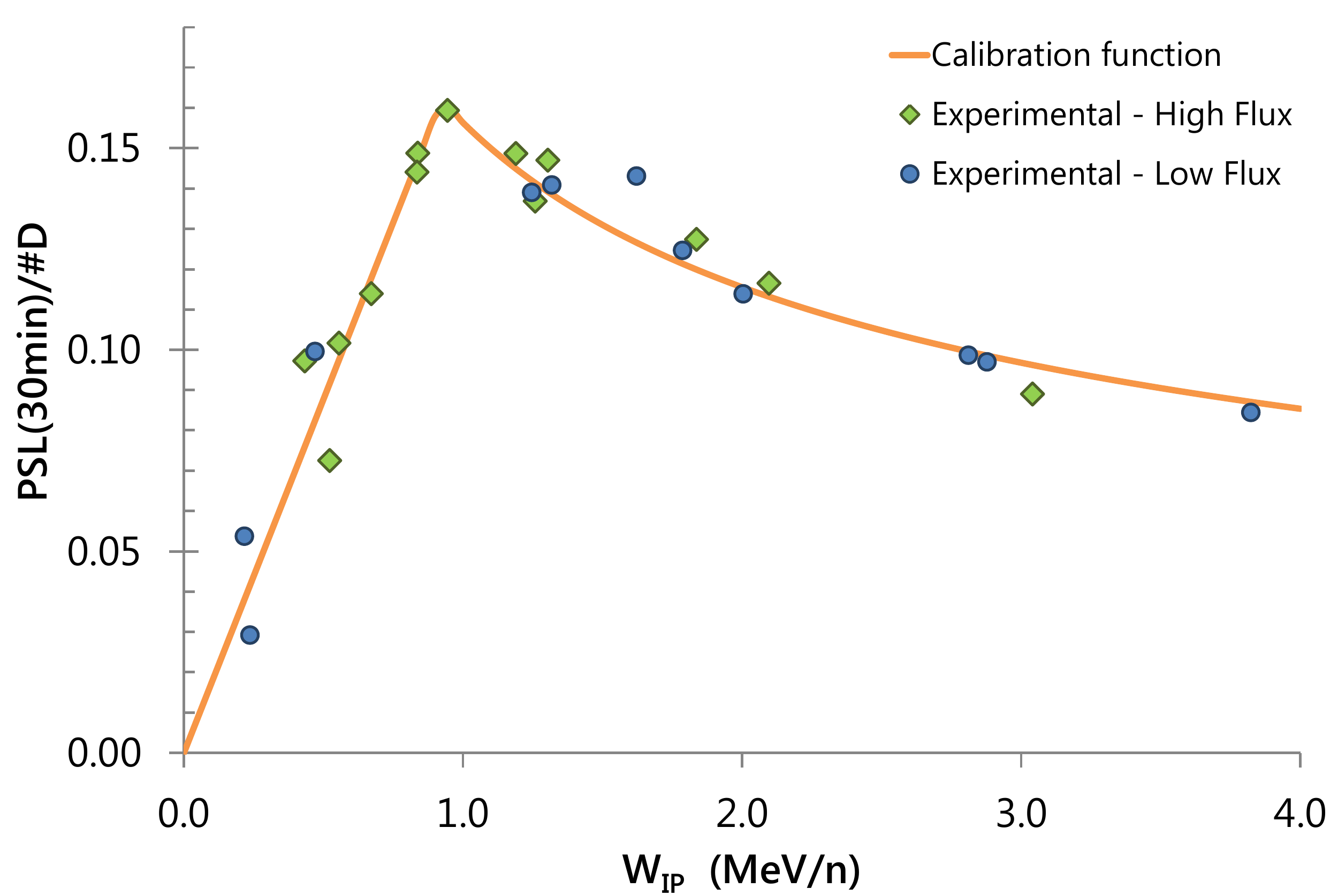}
\caption{Absolute response of BAS-TR image plates to deuterium ions, correlating the PSL on the IP per incident D$^+$ ion for different D$^+$ energies on the IP. The data points are divided into two groups depending on the PSL signal on the IP, which shows the validity of the calibration over a range of ion flux from $\sim\mbox{0.01 ions/}\um^2$ to a few ions/$\um^2$ on the detector.}
\label{fig:Calibration}
\end{figure}

Employing the methods discussed in the subsections A-C, the PSL$_{30}$ per incident particle was obtained for several different D$^+$ energies at the IP, as shown in Fig.~\ref{fig:Calibration}. The calibration data points were obtained over a wide range of PSL values on the IP, which were divided into two groups depending on the signal strength, given by 0.02-0.08 PSL/$\um^2$ and 0.08-0.2 PSL/$\um^2$ in Fig.~\ref{fig:Calibration}, corresponding to low and high particle fluxes, and they are consistent with previous works\cite{freeman2011}. It can be seen how both groups follow the same calibration function, which implies linear response of the Image Plate with respect to the flux for the signal levels in our experiment. In terms of the ion energy, the PSL signal per particle for lower energy ions increases linearly with the incident ion energy, whereas it decreases slowly, following a power law, for higher energy ions. This behaviour of IP response can be crudely explained on the basis of the finite thickness of the IP active layer and the Bragg peak energy deposition process for ions. The empirical calibration function thus obtained by the best fit to the data points is given by Eq.~\ref{eq:Fitting1}.

\begin{equation}\label{eq:Fitting1}
\mbox{PSL}_{30}/\mbox{D}^+=
\begin{cases} 
      0.1754~\mbox{W}_{\mbox{IP}} & \mbox{W}_{\mbox{IP}}<0.931~\mbox{MeV/n} \\
\\
      0.1563~\mbox{W}_{\mbox{IP}}^{-0.4363} & \mbox{W}_{\mbox{IP}}\geq 0.931~\mbox{MeV/n}
\end{cases}
\end{equation}

where, $\mbox{W}_{\mbox{IP}}$ represents the D$^+$ ion energy reaching the IP. Although Fig.~\ref{fig:Calibration} shows the calibration for D$^+$ ions with energies up to 4 MeV/nucleon, Eq.~\ref{eq:Fitting1} can be used to obtain D$^+$ spectra up to significantly higher energies while using DF, as the DF tends to slow down the ions before reaching the IP.

\section{Analysis of a typical deuteron spectrum}\label{sec:Application}

\begin{figure}[h]
\includegraphics[width=0.48\textwidth]{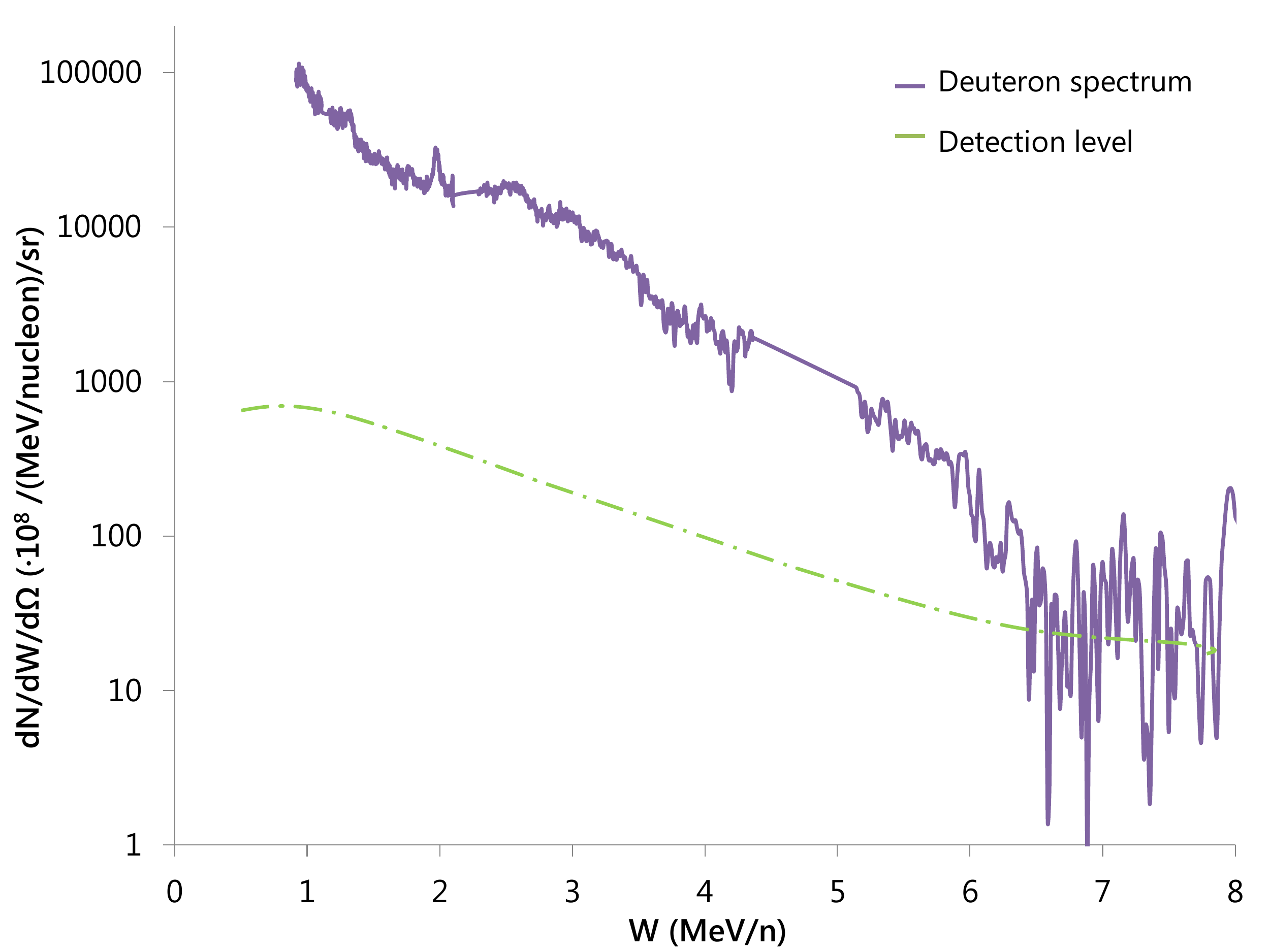}
\caption{Graph shows typical spectra for laser accelerated D$^+$ ions obtained in the experiment, together with the detection threshold of the diagnostics due to noise level of the detector. Dotted lines in the ion spectrum correspond to missing spectral information in the regions between adjacent filters.}
\label{fig:ApplicationLin}
\end{figure}

Once the calibration function had been obtained, it was used to analyse the D$^+$ spectrum obtained in the experiment. As an example, a typical spectrum obtained along the target normal axis has been analysed (Fig. ~\ref{fig:ApplicationLin}). This spectrum corresponds to the interaction of a Petawatt laser pulse with a thin (100~nm) deuterated plastic (CD) target, using the same laser parameters as mentioned in section~\ref{sec:ExperimentalSetup}.

In the case of the data shown in Fig.~\ref{fig:ApplicationLin}, the IPs were scanned $\sim$140 minutes after the irradiation and the signal was then re-normalised to the PSL value at the 30 min reference time (PSL$_{30}$) using Eq. \ref{eq:Fading}. The ion tracks were measured from the parabola on the IP, and background was subtracted by using the PSL values at the regions near the track. For each point on the ion track, the energies of the ions at the image plate were calculated taking into account the energy dispersion of the TPS and the transmission prperties for each filter obtained by SRIM. Finally, the number of D$^+$ ions for each energy was recovered using the calibration function given by the Eq. \ref{eq:Fitting1}.

\section{Conclusions}\label{sec:Conclusions}

In conclusion, a novel technique to obtain the spectrum of ion species over a broad energy range by removing overlapping traces in a Thomson parabola ion spectrometer is demonstrated. This method is based on differential filtering across the full range of energies produced, by which the lightest of the overlapping ions can be discriminated. The technique can also be applied for filtering proton track in a magnetic spectrometers, or, to filter high energy part of an ion spectrum in the region where several ion tracks are merged due to insufficient track separation. The method was employed in a recent experimental campaign to diagnose the deuterium spectrum obtained from the interaction of a high power laser with deuterated plastic targets, which produce a multi-species ion source, with a number of ion species (such as D$^+$, C$^{6+}$, O$^{8+}$, etc.) sharing the same $q/m$ value. In addition to the D$^{+}$ spectrum obtained using DF, the BAS-TR image plate was absolutely calibrated for D$^+$ response against CR-39 detectors over a broad spectrum.

\section*{Acknowledgements}\label{sec:Acknowledgement}
The authors acknowledge funding from EPSRC [EP/J002550/1-Career Acceleration Fellowship held by S. K., EP/L002221/1, EP/E035728/1, EP/K022415/1, EP/J500094/1 and EP/I029206/1], Laserlab Europe (EC-GA 284464), projects ELI (Grant No. CZ.1.05/1.1.00/483/02.0061) and OPVK 3 (Grant No. CZ.1.07/2.3.00/20.0279). Authors also acknowledge the support of the mechanical engineering staff of the Central Laser Facility, STFC, UK.

%

\end{document}